# Digital tools against COVID-19: Framing the ethical challenges and how to address them

Authors: Urs Gasser[1], Marcello Ienca[2], James Scheibner[2], Joanna Sleigh[2], Effy Vayena[2]

[1]Berkman Klein Centre for Internet & Society, Harvard Law School
[2]Health Ethics and Policy Lab, Department of Health Sciences and Technology, ETH Zürich

**Abstract**
Data collection and processing via digital public health technologies are being promoted worldwide by governments and private companies as strategic remedies for mitigating the COVID-19 pandemic and loosening lockdown measures. However, the ethical and legal boundaries of deploying digital tools for disease surveillance and control purposes are unclear, and a rapidly evolving debate has emerged globally around the promises and risks of mobilizing digital tools for public health. To help scientists and policymakers navigate technological and ethical uncertainty, we present a typology of the primary digital public health applications currently in use. Namely: proximity and contact tracing, symptom monitoring, quarantine control, and flow modeling. For each, we discuss context-specific risks, cross-sectional issues, and ethical concerns. Finally, in recognition of the need for practical guidance, we propose a navigation aid for policymakers made up of ten steps for the ethical use of digital public health tools.

**Introduction**
Symptomatic for today's digitally connected society, the collection and use of *data* is presented as a main strategic remedy in response to the COVID-19 pandemic. Across geographies and institutions, public health experts and researchers from diverse fields such as epidemiology, virology, evolutionary biology, and social science have pointed out the broad range of insights that can be gained by collecting, analyzing, and sharing data from diverse digital sources. These sources include data from phone towers, mobile phone apps, Bluetooth connections, surveillance video, social media feeds, smart thermometers, credit card records, wearables, and several other devices and applications. In parallel, Apple and Google, two of the world's largest information technology companies, have unprecedentedly banded together to create a decentralized contact tracing tool to help people determine whether they have been exposed to someone with the SARS-CoV-2 virus.[1]

While the promise of (big) data analysis has been widely acknowledged and governments, and researchers around the globe are rushing to unlock its potential, significant technical limitations have also surfaced. These limitations include the accuracy, granularity, and quality of data that vary greatly across the different data sources, the adequacy of computation safeguards, and interoperability issues and security risks. Simultaneously, important ethical and legal risks and concerns have been identified that accompany digital disease surveillance and prediction.[2] Civil rights organizations, data protection authorities, and emerging scholarship have highlighted the risk of increased digital surveillance following the pandemic.[3] They have emphasized the need to meet baseline conditions such as lawfulness, necessity, and proportionality in data processing, and the need for social justice and fairness to not get lost in the urgency of this crisis.

As numerous public and private sector initiatives aiming to use digital technologies in the fight against COVID-19 continue to emerge, the ensuing debate so far seems to be framed generically in a binary choice between using digital technologies to save lives versus respecting individual privacy and civil liberties. However, a myriad of interdisciplinary research has shown the vital importance of context in managing the societal, legal, and ethical risks of data processing for pandemics.[4–7] In this article, we seek to contribute to the rapidly evolving debate about the promises and risks of digital public health technologies in response to COVID-19. More specifically, we offer a typology of the main applications that are currently in use, and we discuss their respective features, including both applications and context-specific risks, as well as cross-sectional issues and ethical concerns. Finally, we propose a navigation aid for policymakers suggesting steps that should be taken in order to mitigate risks and strike a defensible risk-benefit balance.

Two caveats are important. First, both the typology sketched in this article and the corresponding analysis are likely to evolve as new digital public health technologies designed for COVID-19 are emerging on an almost daily basis. Second, for this article, we consider digital surveillance and contact tracing as part of a broader strategy that is conditioned on large scale testing, universal access to health care, and adequate societal safety nets. Absent these conditions, the use of digital tools is misguided and irresponsible, given the associated risks.

**Typology of digital public health tools**
We identify four main categories of digital public health technologies developed for pandemic management: proximity and contact tracing, symptom monitoring, quarantine control, and flow modeling.
1. Proximity tracing tools measure the spatial proximity between users to track their interaction. Proximity tracing, sometimes also in conjunction with patient reports or other non-digital sources, can identify when users are exposed to a Sars-Cov2 positive individual. Some digital tools combine proximity and contact tracing features. For example, the Singaporean app *TraceTogether* uses Bluetooth connections to log other phones nearby and alerts those who have been close to a Sars-Cov2 positive user. When users have shared proximal space with a Sars-Cov2 positive individual, they are encouraged to self-isolate.[8]
2. Symptom checkers are tools of syndromic surveillance that collect, analyze, interpret, and disseminate health-related data.[9] Using these tools, users report their symptoms, obtain a diagnosis, and get a triage decision. Symptom checkers thus provide a cost-effective way of helping to triage large international and disperse populations of healthcare-seeking patients. Further, the value of digital symptom checkers resides in their enablement of global public health surveillance. Such is exemplified by Spain's *CoronaMadrid* symptom checking app. Using this technology, the government can collaborate with citizens, health professionals, and the private sector to monitor the disease, respond quickly, allocate resources, and generally minimize/ control outbreaks.
3. Quarantine compliance tools involve real-time monitoring of whether symptomatic patients or non-symptomatic individuals are complying with quarantine restrictions. Public health legislation includes requirements for infected or potentially infected individuals to remain isolated from others, so they do not spread the disease further. These technologies can provide a mechanism of controlling that infected individuals remain isolated from other individuals. Examples include Taiwan's *Electronic Fence* that tracks quarantined overseas arrivals using mobile phone data.[10]
4. Flow modeling tools, otherwise known as mobility reports, quantify, and track people's movements in specified geographic regions. Typically, these tools rely on aggregated, anonymized sets of data from the geographic location of users. Flow modeling can provide insights into the effectiveness of response policies (e.g., social distancing or forced quarantine) aimed at combating COVID-19.

This typology allows us to structure these technologies in a four-dimensional model. First, we include the key actors involved in design and implementation (government agencies, academia, private companies, and citizens). Secondly, we assess the different data types being collected, using the classification offered by the GDPR (non-identifying, personal data, and sensitive personal data). Thirdly, our typology includes the different origins of these data, including IP addresses, call site data, GPS data, Bluetooth, and third-party data. Finally, it considers the different types of consent required to collect data, including opt-in consent, opt-out consent, and mandatory use. This four-dimensional model allows us to compare the ethical implications of different types of technological approaches to pandemic management, as demonstrated in the diagram below.

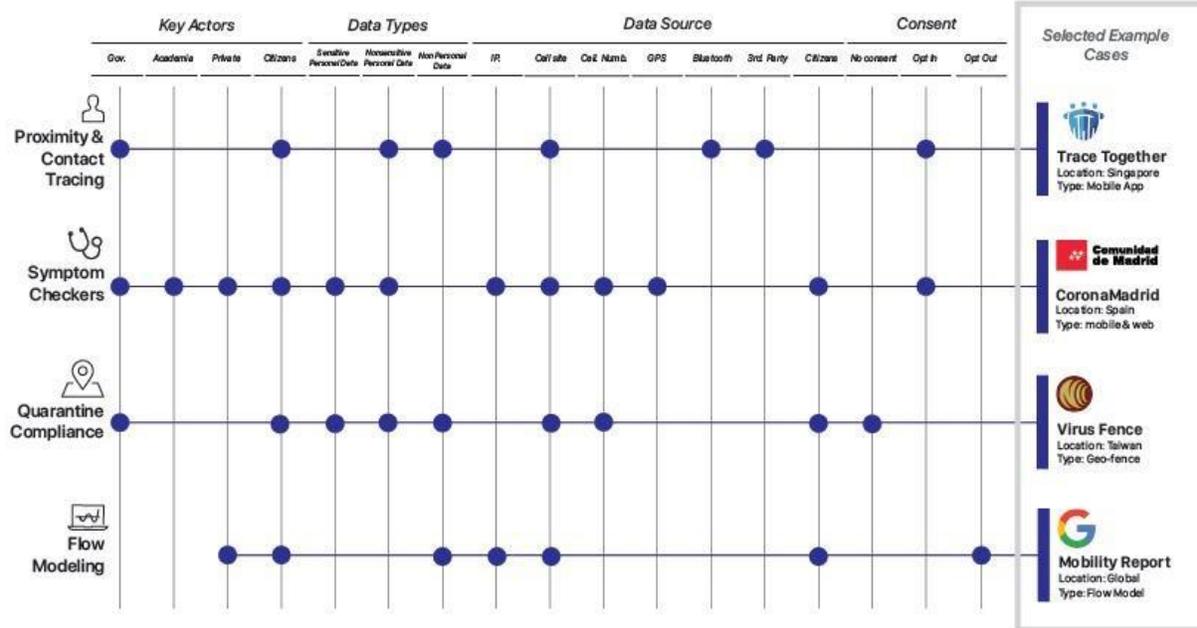

Figure 1- Typology of Digital Public Health Technologies against COVID-19

**Ethical and Legal Challenges**

These four types of digital public health technologies raise both cross-sectional and domain-specific ethical-legal considerations. These considerations are rooted in the basic principles and moral considerations of public health ethics and data ethics.[11,12]

**Scientific validity, accuracy, and data necessity:** Despite widespread enthusiasm about using digital public health technologies to combat epidemics, this use involves inevitable compromise. On the one hand, digital public health technologies can improve the rapidity of pandemic response.[13] On the other hand, low data quality and integrity flowing from technical issues can have an outsized effect on large-scale predictive models.[14] Additionally, the effectiveness of digital public health technologies tools for contact tracing depends upon uptake, which will vary according to location, the existence of other measures, and disease prevalence. Uncertainty about scientific validity and efficacy can make assessing the proportionality and risk of proposed measures more challenging.[15] Subsequently, measures based on such models may be disproportionate, negatively affecting individuals and populations without generating significant benefits. While several global actors are independently pursuing digital public health strategies, typically at the country-level, it is critical to ensure the interoperability of such digital systems and enable efficient, harmonized and secure cross-national data sharing.[16] This pandemic is a global challenge, hence cannot be tackled only locally and requires new cooperative approaches.

**Privacy:** All digital public health tools impinge upon individual privacy by requiring some degree of access to information about the health status, behavior, or location of individuals.[17] However, privacy risks vary depending on the purpose and data types used by a digital tool. Digital tools for measuring relative spatial proximity among phone users are, *ceteris paribus*, less privacy-invasive than personal contact tracing or quarantine enforcement apps. Likewise, tools using aggregate mobile phone tower data are, on average, less privacy-invasive than tools based on GPS data and sensor tracking for individual users.[18] The use of more granular and specific types of data can increase the risk of downstream reidentification of individuals or groups. Further, with the vast amount of data being gathered, public health agencies and app developers must prevent downstream reidentification through data linkage.[19] It is also vital to understand that privacy risks can change and accumulate over time, which highlights the importance of strong legislative protection. . In the European Union (EU), several regulatory instruments offer varying levels of safeguards for the right to privacy, and data protection. These include the General Data Protection Regulation (GDPR), the e-Privacy Directive and the Charter of Fundamental Rights (CFR). Likewise, at the Council of Europe level, the European Charter of Human Rights (ECHR) guarantees the right to "private life." However, these regulations set forth the circumstances where these rights can be abridged, including during a public health crisis. Further, any digital public health technologies abridging these rights must be proportionate to the aims sought. In other words, this abridgment must lead to faster restoration of other rights and freedoms suspended due to lockdown policies (e.g., freedom of movement and freedom of assembly).[20]

**Consent & voluntariness:** Digital public health technologies have the potential to undermine not only privacy but also personal autonomy. Specifically, smartphone apps often include permissions to collect data beyond the stated purpose of the app. These data handling practices might strip people of their ability to consent to being tracked or having their information shared,[21] depending on their purpose, mode of data collection, and data source. For example, in order to work properly, proximity tracking apps based on Bluetooth need to require or encourage users to keep their Bluetooth turned on at all times, creating additional risks. These approaches to data collection must respect autonomy, such as by ensuring strategies are in place to update the user regularly.[22] Finally, mandating quarantine apps or technologies for infectious individuals or their contacts raises the most severe questions of justifiable coercion. On the one hand, the effectiveness of quarantine might be undermined if it remains voluntary rather than mandatory. On the other hand, some government activity (such as the Polish government creating shadow profiles for returning citizens as part of a quarantine app) may constitute an overreach on autonomy.[23]

**Discrimination:** Along with the risk of reidentification and infringement of personal autonomy, digital public health technologies also carry an inherent risk of discrimination. Specifically, these technologies can be used to collect large amounts of data about entire populations. This data can include race, ethnic group, gender, political affiliation, and socio-economic status, which in turn can be used to stratify populations by demographics. Many of these demographics are sensitive and not necessarily related to a person's health, and may lead to stigmatization.[24] Further, information such as racial demographics might lead to a surge in discrimination, as seen by a rise in attacks on people of South East Asian descent in the COVID-19 crisis. Therefore, safeguards must exist for any digital public health technologies to prevent 'the predictable from becoming exploitable'.[19]

**Repurposing:** There is a risk that digital tools could also be applied to other forms of surveillance as well as used for legitimate public health purposes (namely tracking and monitoring COVID-19 patients), but also applied to other forms of surveillance. For example, one NYT report investigated Health Code, an Alibaba-backed government-run app that supports decisions about who should be quarantined for COVID-19 in China. The report discovered that the app also appears to share information with the police.[25] Further, the UK and the US have developed biosurveillance programs that share the characteristics of both pandemic response and counter-terrorist programs.[26] Therefore, it is crucial to distinguish digital public health technologies that allow this third-party sharing of information for non-health-related purposes from those which do not.

**Expiration:** Pandemics are a rare situation where democratic governments can take unchecked executive action decisions for the collective good of their population. These include actions that might be in contravention of political due process or individual human rights. If prolonged, these actions can deprive citizens of their rights, with no guarantee these rights will be restored after the end of the crisis. The USA Patriot Act, promulgated after the September 11 terrorist attacks in the US, is a good example of how democratic liberties might be ceded after an emergency. Likewise, there was an outcry after the regime of Viktor Orban in Hungary instituted powers by decree to fight the COVID-19 pandemic without an expiration date. Therefore, heightened surveillance empowered by digital public health technologies should not continue after the

COVID-19 pandemic has ended. Further, such programs should clarify upfront the duration clearly what data they are collecting and the time limit on how long they will hold the information.[27]

**Digital inequality:** Digital technology, particularly mobile technology, is increasingly widespread globally but unevenly distributed. In 2019, two-thirds of the world's population did not own a smartphone technology, and one third did not own any mobile phone. Smartphone ownership disparities are particularly noticeable in emerging economies. For instance, in India, the world's second-most populous country accounting for over 17% of the global population, only 24% of adults report owning a smartphone. Even in advanced economies with high smartphone ownership rates, not all age cohorts are catching up with digital tools. In 2018, most citizens of Japan, Italy, and Canada over the age of 50 did not own a smartphone.[28] Any digital public health technology solution which relies on mobile phones excludes those without access to these technologies for geographic, economic, or demographic reasons. Digital public health technologies may be less helpful for both people in low- and middle-income countries and older people. If not complemented with non-digital strategies, this latter risk might exacerbate health inequalities.

**Public Benefit:** Underpinning all of the ethical and legal challenges mentioned so far regarding pandemic management is the question of public benefit. For the use of digital public health tools to be proportionate to any rights impinged, there must be a clear public benefit from their use. These benefits can include potential forecasting outbreaks,[29] preventing or reducing new infections, improving the efficiency of social care and vaccine development, and improving how information is communicated to citizens.[24] However, to offset the risks and infringement of individual rights, there must be a framework for deciding what public benefit is appropriate. In this regard, Laurie suggests the test of 'reasonable benefit' in the context of data sharing for pandemic response.[30] Nevertheless, what is reasonable for a digital public health technology will very much depend on other risks associated with that technology.

**Mapping ethical and legal challenges**

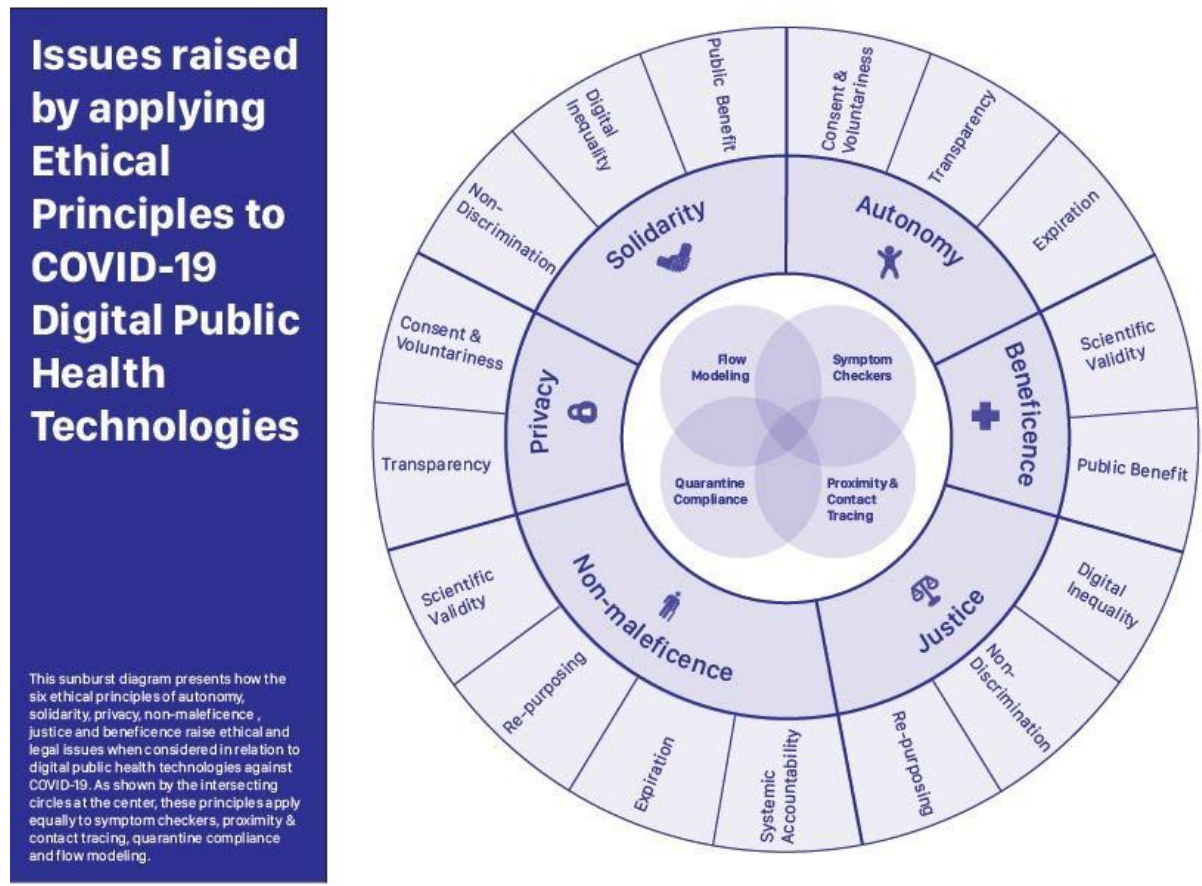

Figure 2- Sunburst diagram mapping the ethical & legal issues raised by applying ethical principles to COVID-19 Digital Public Health Technologies

**Ethical use of digital public health tools: A Navigation Aid**

Decision-makers who seek to embrace any of the emerging COVID-19 digital public health technologies need to address their ethical and legal issues. Further, these decision-makers must put safeguards into place to avoid harm and manage the remaining risks. Best practices still have to emerge for COVID-19 digital public health technologies. However, given the unique circumstances of the current pandemic, procedural guidance - a navigation aid in the form of an iterative set of steps to work through - can be useful. In addition, frameworks for ethical data uses that have articulated ethical values and several procedural principles such as adaptivity, flexibility, reflexivity, transparency, accountability, responsiveness.[31,32] Using these frameworks, we propose the following aid. This navigation aid aims to assist decision-makers in ensuring that digital public health tools are used throughout their lifecycle in a legally and ethically responsible way.[10]

- *Establish guiding ethical principles:* In addition to ensuring compliance with fundamental rights and applicable legal norms, establish clarity with respect to value commitments, red lines, and guiding principles that will help to navigate tensions or conflicts between values when embracing digital technology against the fight of COVID-19.[33–36]

- *Distinguish tools from purpose:* Define specific objectives within the containment/mitigation strategy. Only then consider the different data sources and means to collect, use, and otherwise process them. Operate within the realm of established laws, regulations, and best practices without recourse to emergency regulation or measures.
- *Avoid lock-in and path dependency*: Consider the range of tools, techniques, etc. and data governance models available once the questions and goals have been determined. Understand what the different instruments and models can and can't do, what their promise and limitations are, and use the above list of the technical, legal, and ethical core issues as evaluation criteria.
- *Conduct risk assessments*: For each intended purpose, context, instrument, and model, engage in a robust and systematic risk assessment process even when pressed for time; well-established practices such as human rights impact assessment and privacy risk impact assessment should lead the way, even if need to be modified.[37] Do not limit the assessment to a question of compliance; apply a holistic ethics perspective taking into account the substantive issues listed in Part 2 of this article.
- *Plan preemptively*: Consider the full lifecycle of data and systems and include both online and offline effects,[38] when conducting risk assessments and determining appropriate safeguards. Special consideration is due to context shifts over time and unintended consequences, second and third-order effects, and the like. For example, while a proximity tracing tool might be privacy-preserving, identification might occur downstream when the person has to be isolated or quarantined.
- *Embrace privacy "by design" and "by default" approaches*:[39] In terms of safeguards, consider and combine the most effective legal, organizational, and technical measures, including advanced statistical and computational safeguards to manage privacy and data protection risks and address ethical issues. Adopt "privacy by design and by default" principles from the outset, and build out additional protective layers over time.
- *Assemble the right team*: The technical, organizational, legal, ethical, public health and other challenges that need to be managed when using digital tools in response to COVID-19 are complex and require an interdisciplinary team. Ensure to assemble a team from diverse backgrounds, with diverse experiences, and high integrity and participate in communities of practice.[40]
- *Communicate proactively and continuously*: Transparency in the form of provocative communication with the key stakeholders - and where possible active consultation and participation with the public - is essential and needs to be an integral part of the process from beginning to end. Form communities of practice and also learn from the experiences of partners and collaborators.
- *Create systemic accountability*: Establish mechanisms to monitor how things work in practice, not only as a matter of compliance but also in terms of unanticipated ethical ramifications. Leverage existing institutional arrangements and processes and aim for independent, external oversight that brings together expertise from different fields to oversee the use of ethical health tools, develop stopping rules, conduct period reviews, etc. Following the systemic oversight framework, this accountability architecture should be sufficiently adaptive, flexible, inclusive, and reflexive to account for the ever-evolving digital public health ecosystem.[32]
- *Keep records and capture learnings*: Throughout these steps, documentation is essential, both of the risk assessment itself as well as the safeguards and accountability mechanisms that have been taken to mitigate remaining risks, and serves as a basis for continued learning, also from mistakes.

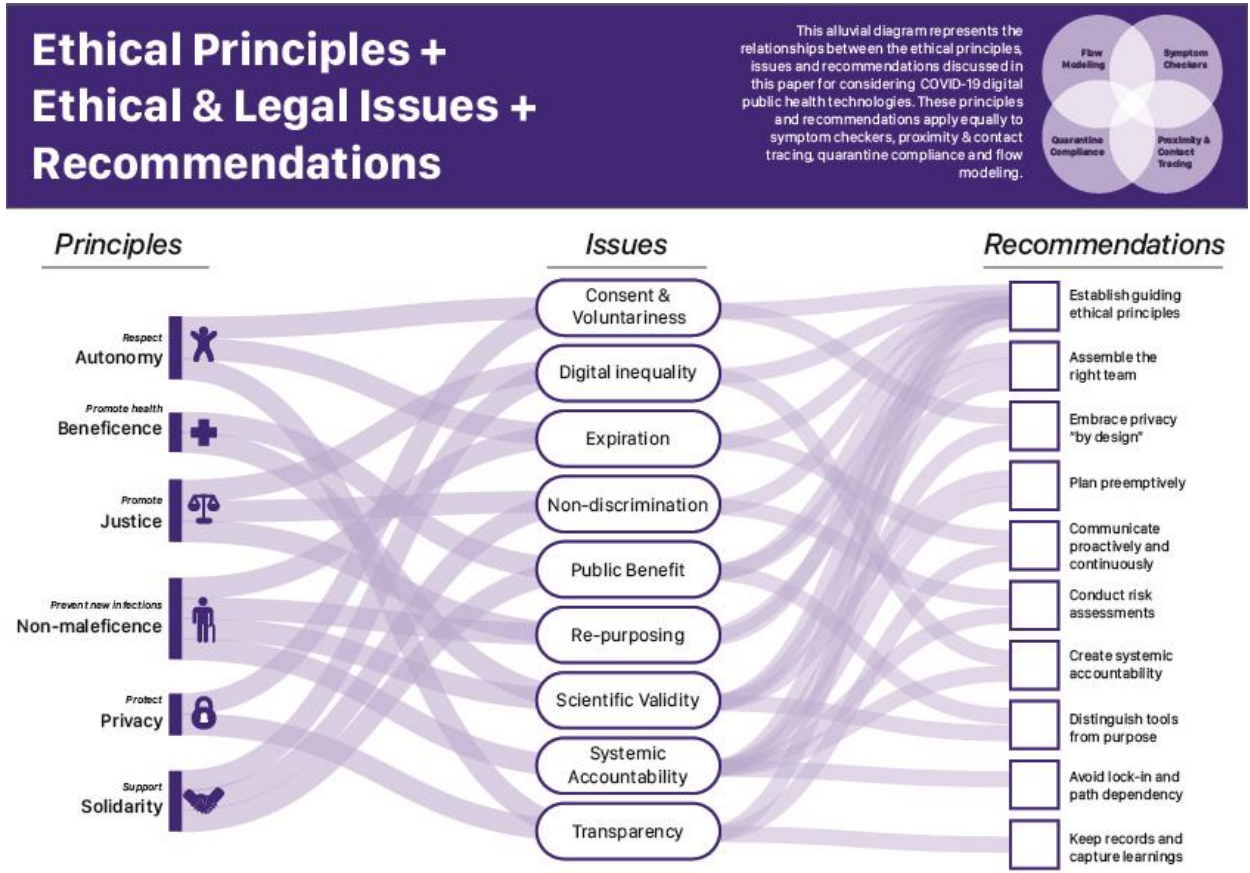

Figure 3: Alluvial Diagram representing the relationship between ethical principles, ethical & legal Issues, & recommendations

In the wake of the COVID-19 pandemic, there has been a surge in interest for the use of digital public health technologies for pandemic management. However, these tools must be guaranteed as ethically compliant to ensure widespread public trust and uptake. In this article, we construct a typology of the main types of digital public health technologies used to fight the COVID-19 pandemic, as well as their ethical impacts. We conclude by providing a navigation aid for identifying and resolving these ethical impacts where possible.

# References

1. Apple and Google partner on COVID-19 contact tracing technology. Google. 2020; published online April 10. https://blog.google/inside-google/company-announcements/apple-and-google-partner-covid-19-contact-tracing-technology (accessed April 14, 2020).

2. Ienca M, Vayena E. On the responsible use of digital data to tackle the COVID-19 pandemic. *Nature Medicine* 2020; **26**: 463–4. DOI:10.1038/s41591-020-0832-5.

3. COVID-19, digital surveillance and the threat to your rights. Amnesty International. 2020; published online April 3. https://www.amnesty.org/en/latest/news/2020/04/covid-19-surveillance-threat-to-your-rights/. (accessed April 17, 2020).

4. Gostin L. Public Health Strategies for Pandemic Influenza. *JAMA* 2006; **295**: 1700.

5. Martin R, Conseil A, Longstaff A, *et al.* Pandemic influenza control in Europe and the constraints resulting from incoherent public health laws. *BMC Public Health* 2010; **10**.

6. Laurie GT, Hunter KG. Mapping, Assessing and Improving Legal Preparedness for Pandemic Flu in the United Kingdom. *Medical Law International* 2009; **10**: 101–37.

7. Aghaizu A, Elam G, Ncube F, et al. Preventing the next 'SARS' - European healthcare workers' attitudes towards monitoring their health for the surveillance of newly emerging infections: qualitative study. BMC Public Health 2011; **11**. DOI:10.1186/1471-2458-11-541.

8. Cho H, Ippolito D, Yu YW. Contact Tracing Mobile Apps for COVID-19: Privacy Considerations and Related Trade-offs. *ArXiv* 2020; published online Mar 30. arXiv: 2003.11511v2 (preprint).

9. Berry AC. Online Symptom Checker Applications: Syndromic Surveillance for International Health. *Ochsner Journal* 2018; **18**: 298–9.

10. Wang CJ, Ng CY, Brook RH. Response to COVID-19 in Taiwan. JAMA 2020; **323**: 1341.

11. Coughlin SS. How Many Principles for Public Health Ethics? *The Open Public Health Journal* 2008; **1**: 8–16.

12. Childress JF, Faden RR, Gaare RD, et al. Public Health Ethics: Mapping the Terrain. *The Journal of Law, Medicine & Ethics* 2002; **30**: 170–8.

13. Yan SJ, Chughtai AA, Macintyre CR. Utility and potential of rapid epidemic intelligence from internet-based sources. *International Journal of Infectious Diseases* 2017; **63**: 77–87.

14. Ferretti L, Wymant C, Kendall M, *et al.* Quantifying SARS-CoV-2 transmission suggests epidemic control with digital contact tracing. *Science* 2020. DOI:10.1126/science.abb6936.

15. Servick K. Cellphone tracking could help stem the spread of coronavirus. Is privacy the price? *Science* 2020; published online March 22. https://www.sciencemag.org/news/2020/03/cellphone-tracking-could-help-stem-spread-coronavirus-privacy-price (accessed April 9, 2020).

16. Kelion L. BBC News. 2020; published online April 12. https://www.bbc.com/news/technology-52263244 (accessed April 13, 2020).

17. Lee LM, Heilig CM, White A. Ethical Justification for Conducting Public Health Surveillance Without Patient Consent. American Journal of Public Health 2012; **102**: 38–44.

18. Wiewiórowski WR. Monitoring the spread of COVID-19, European Data Protection Supervisor. 2020; published online March 25. https://edps.europa.eu/sites/edp/files/publication/20-03-25_edps_comments_concerning_covid-19_monitoring_of_spread_en.pdf.

19. Garattini C, Raffle J, Aisyah DN, Sartain F, Kozlakidis Z. Big Data Analytics, Infectious Diseases and Associated Ethical Impacts. *Philosophy & Technology* 2017; **32**: 69–85.

20. Kolfschooten HV. EU Coordination of Serious Cross-Border Threats to Health: The Implications for Protection of Informed Consent in National Pandemic Policies. *European Journal of Risk Regulation* 2019; **10**: 635–51.

21. Mitchell SSD. 'Warning! You're entering a sick zone'. *Online Information Review* 2019; **43**: 1046–62.

22. Geneviève LD, Martani A, Wangmo T, et al. Participatory Disease Surveillance Systems:

Ethical Framework. *Journal of Medical Internet Research* 2019; **21**: e12273.

23. Hamilton IA. Poland made an app that forces coronavirus patients to take regular selfies to prove they're indoors or face a police visit. Business Insider. 2020; published online March 23. https://www.businessinsider.com/poland-app-coronavirus-patients-mandaotory-selfie-2020-3 (accessed April 9, 2020).

24. Quinn P. Crisis Communication in Public Health Emergencies: The Limits of 'Legal Control' and the Risks for Harmful Outcomes in a Digital Age. *Life Sciences, Society and Policy* 2018; **14**. DOI:10.1186/s40504-018-0067-0.

25. Mozur P. In Coronavirus Fight, China Gives Citizens a Color Code, With Red Flags. New York Times. 2020; published online March 1. https://www.nytimes.com/2020/03/01/business/china-coronavirus-surveillance.html (accessed April 9, 2020).

26. Parry B. Domesticating biosurveillance: 'Containment' and the politics of bioinformation. *Health & Place* 2012; **18**: 718–25.

27. Ng ES, Tambyah PA. The ethics of responding to a novel pandemic. *Ann Acad Med Singap* 2011; **40:** 30-5.

28. Silver L. Pew Research Center's Global Attitudes Project. Pew Research Center's Global Attitudes Project. 2019; published online Feb 5. https://www.pewresearch.org/global/2019/02/05/smartphone-ownership-is-growing-rapidly-around-the-world-but-not-always-equally/ (accessed April 9, 2020).

29. Roberts SL. Big Data, Algorithmic Governmentality and the Regulation of Pandemic Risk. *European Journal of Risk Regulation* 2019; **10**: 94–115.

30. Laurie GT. Cross-Sectoral Big Data. *Asian Bioethics Review* 2019; **11**: 327–39.

31. Xafis V, Schaefer GO, Labude MK, et al. An Ethics Framework for Big Data in Health and Research. *Asian Bioethics Review* 2019; **11**: 227–54.

32. Vayena E, Blasimme A. Health Research with Big Data: Time for Systemic Oversight. The

*Journal of Law, Medicine & Ethics* 2018; **46**: 119–29.

33. Olbrechts A. Statement on the processing of personal data in the context of the COVID-19 outbreak. European Data Protection Board - European Commission. 2020; published online March 20. https://edpb.europa.eu/our-work-tools/our-documents/other/statement-processing-personal-data-context-covid-19-outbreak_en (accessed April 11, 2020).

34. The-GSMA-COVID-19-Privacy-Guidelines. GSMA |Public Policy. 2020; published online April. https://www.gsma.com/publicpolicy/resources/covid-19-privacy-guidelines/the-gsma-covid-19-privacy-guidelines. (accessed April 17, 2020).

35. Joint Civil Society Statement: States use of digital surveillance technologies to fight pandemic must respect human rights. Human Rights Watch. 2020; published online April 2. https://www.hrw.org/news/2020/04/02/joint-civil-society-statement-states-use-digital-surveillance-technologies-fight (accessed April 11, 2020).

36. NGO and expert statement to the OECD Secretary General on COVID-19, privacy and fundamental rights. Association for Progressive Communications. 2020; published online April 1. https://www.apc.org/en/pubs/ngo-and-expert-statement-oecd-secretary-general-covid-19-privacy-and-fundamental-rights (accessed April 11, 2020).

37. Allison-Hope D, Vaughn J. BSR. BSR. 2020; published online March 30. https://www.bsr.org/en/our-insights/blog-view/covid-19-a-rapid-human-rights-due-diligence-tool-for-companies (accessed April 11, 2020).

38. Altman M, Wood A, O'Brien DR, Gasser U. Practical approaches to big data privacy over time. *International Data Privacy Law* 2018; **8**: 29–51.

39. Greenwood D, Nadeau G, Tsormpatzoudi P, Wilson B, Saviano J, Pentland A 'Sandy'. COVID-19 Contact Tracing Privacy Principles. *MIT Comput Law Rep* 2020; published online April 6. https://law.mit.edu/pub/covid19contacttracingprivacyprinciples (accessed April 11, 2020).

40. Ingram D, Ward J. Behind the global efforts to make a privacy-first coronavirus tracking app.

NBCNews.com. 2020; published online April 7. https://www.nbcnews.com/tech/tech-news/behind-global-efforts-make-privacy-first-coronavirus-tracking-app-n1177871 (accessed April 11, 2020).